# KAON ELECTROPRODUCTION OFF PROTON[*]

## N. GHAHRAMANY[**] AND M. GHANAATIAN

Department of Physics, Shiraz University, Shiraz, I. R. of Iran, 71454
ghahramany@physics.susc.ac.ir

**Abstract** – The elementary reaction of kaon exclusive electroproduction on protons has been studied in a broad kinematical range. Data for the calculation of the form factor of the kaon have been taken at different values of the invariant center of mass energy W in the range W=1.8, 1.85, 1.98, 2.08 (GeV), for one value of the transferred 4-momentum $Q^2$=2.35 $(GeV/c)^2$. In this analysis, we calculated $\sigma_L$ and $\sigma_T$ by using the Rosenbluth separation. Then the electromagnetic kaon form factor was calculated by Chew-Low extrapolation.

**Keywords –** Electroproduction, kaon, form factor, rosenbluth separation

## 1. INTRODUCTION

The study of nuclear physics provides a description of the forces, reactions, and internal structures of hadronic matters. Understanding nuclear physics is the key to understanding the universe, as matter comes from nuclei and nuclear reactions. The nucleus, as a collection of baryons in close proximity provides an ideal microscopic laboratory for testing the structure of fundamental interactions. One of the most intriguing areas of intermediate energy nuclear physics (i.e., energies of a few GeV) is the interface between a hadron description and a quark description of nuclei and subnuclear processes. A hadron description such as Quantum Hadrodynamics (QHD) considers a multitude of mesons and baryons such as pions, kaons, protons, neutrons, Δ's, Λ's, and Σ's to be the fundamental objects which interact with one another. Generally, hadrons are divided into two classes of particles: baryons and mesons. The nucleus of an atom contains baryons, which are a complex system of quarks and gluons. Baryons have half-integral spin and participate in strong, electromagnetic, and weak interactions. Mesons have integral spin and masses generally intermediate between leptons and baryons. Baryons are composed of quark triplets ($qqq$), while mesons are composed of quark-antiquark ($q\bar{q}$) pairs.

Baryons are now understood as complicated many body systems, comprised of quarks and gluons whose interactions are described by quantum chromodynamics (QCD). The confinement of quarks and gluons in the interior of the hadrons and the small coupling constant of quarks at very small distances (asymptotic freedom) are two of the most important properties described by QCD. Although QCD provides the most complete description of quark and gluon interaction, it is most difficult, and sometimes impossible, to solve analytically.

The use of electromagnetic probes to study the nuclear structure, and nucleon structure with higher energies, is ideal because the electromagnetic interaction is governed by quantum

---

electrodynamic (QED). Electron interactions are well understood and described accurately by QED. There is minimal disturbance to the target because electromagnetic interactions are weak compared to the strong interactions involved with using hadronic probes. This weak interaction allows the electron to probe the entire nucleus with minimal disturbance to the target.

## 2. KAON ELECTROPRODUCTION

After its discovery in the 1940's, strange physics was a very active field of study for about two decades. Despite some early successes, the field of electromagnetic production of strangeness was gradually abandoned in the mid to late 1970s, mainly due to a lack of adequate experimental facilities and an apparently complicated reaction mechanism. As a direct consequence of this lack of activity in the field, the experimental data is very scarce, and for the most part, plagued by large statistical and systematic uncertainties.

In deep inelastic scattering experiments three different event classes can be distinguished [1-3]. In each subsequently listed event class the requirements are more restrictive.
- Inclusive events, when only the scattered lepton is identified.
- Semi-inclusive events, when also one or more of the produced hadrons are detected.
- Exclusive events, corresponding to scattering processes in which the target nucleon remains intact or ends up in a lowly excited baryon resonance state; experimentally, it is required that both the scattered lepton and all produced particles are detected (or indirectly reconstructed).

The reaction $e + A \rightarrow e' + K^+ + Y + (A-1)$ forms the basic reaction for the hadron electroproduction of a nuclear target [1, 4-8]. In the above reaction, Y is either a $\Lambda$, $\Sigma^0$ or $\Sigma^-$ hyperon and (A-1) represents the remaining nucleons. For A=1 and assuming one-photon exchange, this can be written as

$$e + p \rightarrow e' + K^+ + \Lambda$$

$$e + p \rightarrow e' + K^+ + \Sigma$$

$$e + n \rightarrow e' + K^+ + \Sigma^-$$

The definition of the kinematical variables (Fig. 1) are given for the reaction

$$e + p \rightarrow e' + K^+ + Y(\Lambda / \Sigma^0)$$

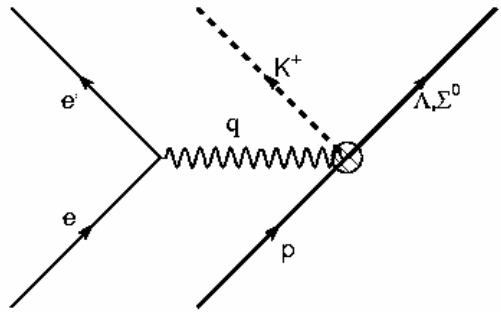

Fig. 1. Feynman diagram representation of kaon electroproduction
in the one-photon exchange approximation

The four-momentum involved are:

- $e = (E, \vec{p}_e)$ for the incident electron.
- $e' = (E', \vec{p'}_e)$ for the scattered electron.
- $p = (M_p, \vec{0})$ for the target nucleon.
- $k = (E_K, \vec{p}_K)$ for the produced kaon.
- $Y = (E_Y, \vec{p}_Y)$ for the unobserved residual system.

A few Lorentz invariants and other kinematic variables are defined below as:

- $q = (\nu, \vec{q})$ is the four-momentum transfer.
- $\nu = E - E'$ is the energy of the virtual photon.
- $\vec{q} = (\vec{p}_e - \vec{p'}_e)$ is the three-vector-momentum of the virtual photon.
- $q^2 = (e - e')^2 = -4EE'\sin^2(\theta_e/2) = -Q^2$ is the square of the four-momentum transfer carried by the virtual photon.
- $W^2 = (q + p)^2 = M_p^2 + 2M_p\nu - Q^2$ is the mass squared of the system recoiling against the electron (i.e. the photon-proton system).
- $t = (q - k)^2 = q^2 + m_K^2 - 2qk$ is the Mandelstam variable.
- $s = (e + p)^2 = M_p^2 + 2M_p E$ is the Mandelstam variable.
- $u = (k - p)^2$ is the Mandelstam variable.
- $x = \dfrac{Q^2}{2M_p\nu}$ is the Bjorken scaling variable (interpreted in the quark parton model, as the fraction of the target nucleon's momentum carried by the struck quark).

The coincidence cross-section can be written in terms of a five-fold differential cross section

$$\frac{d^5\sigma}{dE'd\Omega'_e d\Omega_K}.$$

The electroproduction cross section is often more convenient to express the results in terms of the CM cross-section. This way one can make direct comparisons with previous (if any) or similar (say we are interested in comparing kaon and pion electroproduction) measurements [1, 5, 9].

One can express the Laboratory cross-section in terms of the CM cross-section via:

$$\frac{d^5\sigma}{dE'd\Omega'_e d\Omega_K} = \Gamma \frac{d\cos\theta^*}{d\cos\theta} \frac{d^2\sigma}{d\Omega^*_K}$$

where $\Gamma$ is the virtual-photon flux factor:

$$\Gamma = \frac{\alpha}{2\pi^2} \frac{(W^2 - M_p^2)}{2M_p^2} \frac{E'}{E} \frac{1}{Q^2} \frac{1}{(1-\varepsilon)}$$

and $\dfrac{d\cos\theta^*}{d\cos\theta}$ is simply the Jacobian between the CM ($\theta^*$) and the Laboratory ($\theta$) angle between the virtual photon and the kaon. $d\Omega'_e = d\cos\theta d\varphi$ is the electron Lab frame solid angle and $d\Omega_K = d\cos\theta_K d\varphi$ is the kaon Lab frame solid angle and $d\Omega^*_K = d\cos\theta^*_K d\varphi$ is the kaon

Center-of-momentum frame (CM) solid angle. $\frac{d^2\sigma}{d\Omega_K^*}$ represents the CM (sometimes called reduced) cross-section and ε is the virtual photon polarization:

$$\varepsilon = \frac{1}{1+2(1+\frac{\nu^2}{Q^2})\tan^2(\theta_e/2)}$$

Finally, in the center-of-mass frame, the cross section for the kaon electroproduction can be expressed as [1, 9-11]:

$$\frac{d^2\sigma}{d\Omega_K^*} = \left(\frac{d^2\sigma_T}{d\Omega_K^*}\right) + \varepsilon\left(\frac{d^2\sigma_L}{d\Omega_K^*}\right) + \varepsilon\left(\frac{d^2\sigma_{TT}}{d\Omega_K^*}\right)\cos 2\varphi + \sqrt{2\varepsilon(\varepsilon+1)}\left(\frac{d^2\sigma_{LT}}{d\Omega_K^*}\right)\cos\varphi$$

or more compactly,

$$\frac{d^2\sigma}{d\Omega_K^*} = \sigma_T + \varepsilon\sigma_L + \varepsilon\cos(2\varphi)\sigma_{TT} + \sqrt{2\varepsilon(\varepsilon+1)}\cos(\varphi)\sigma_{LT}$$

where:

- $\sigma_T$ is the cross section due to transversely polarized virtual photons.
- $\sigma_L$ is the cross section due to longitudinally polarized virtual photons.
- $\sigma_{LT}$ is the cross section due to interference between transversely polarized and longitudinally polarized virtual photons.
- $\sigma_{TT}$ is the cross section due to interference between the two different states of transversely polarized virtual photons.
- ε is the virtual photon polarization.
- φ is the azimuthal angle between the scattering and production planes.

The quantities $\sigma_T$, $\sigma_L$, $\sigma_{TT}$ and $\sigma_{LT}$ completely characterize the dependence of the cross-section on the nucleon (nucleus). Generally they are functions of the kinematical variables $Q^2$, W, and t.

The separation of $\sigma_T$ and $\sigma_L$ is possible via the so-called Rosenbluth technique [1, 10]. If one integrates the cross section over all $\varphi \in (0, 2\pi)$, the interference terms vanish, leaving only the combined contributions from the transverse and longitudinal cross sections, $\sigma_T + \varepsilon\sigma_L$. Then, by measuring the cross section at several values of the virtual photon polarization, ε, the cross sections $\sigma_T$ and $\sigma_L$ can be separated.

A least square fitting routine was then used to fit a linear dependence of ($\sigma_T + \varepsilon\sigma_L$) vs. ε. The results of the fit were used to extract the separated values of $\sigma_T$ (the ε = 0 intercept of the fitted line) and $\sigma_L$ (the slope of the fitted line) for each $Q^2$.

## 3. THE ELECTROMAGNETIC KAON FORM FACTOR

We can study the t-dependence of the electroproduction process to isolate the t-channel exchange contribution and extract the kaon electromagnetic form factor [12-14].

In the t-channel the virtual photon directly couples to the kaon. Only longitudinally polarized photons will scatter in the forward direction from the kaon (or mesonic current of the nucleon) in the t-channel; and the kaon form factor is thus extracted from the longitudinal component. This contribution is enhanced when the variable |t| is at a minimum, which depends on the squared virtual photon mass. Most of the physics discussed here will seek to separate out the t-channel $K^+$ production mechanism from the other possibilities because this is a relevant mode that involves the kaon electromagnetic form factor, the Born term in the t-channel. The s- and u-channel Born terms are shown in Fig. 2. Higher-order contributions are not indicated, but their contributions are worth consideration.

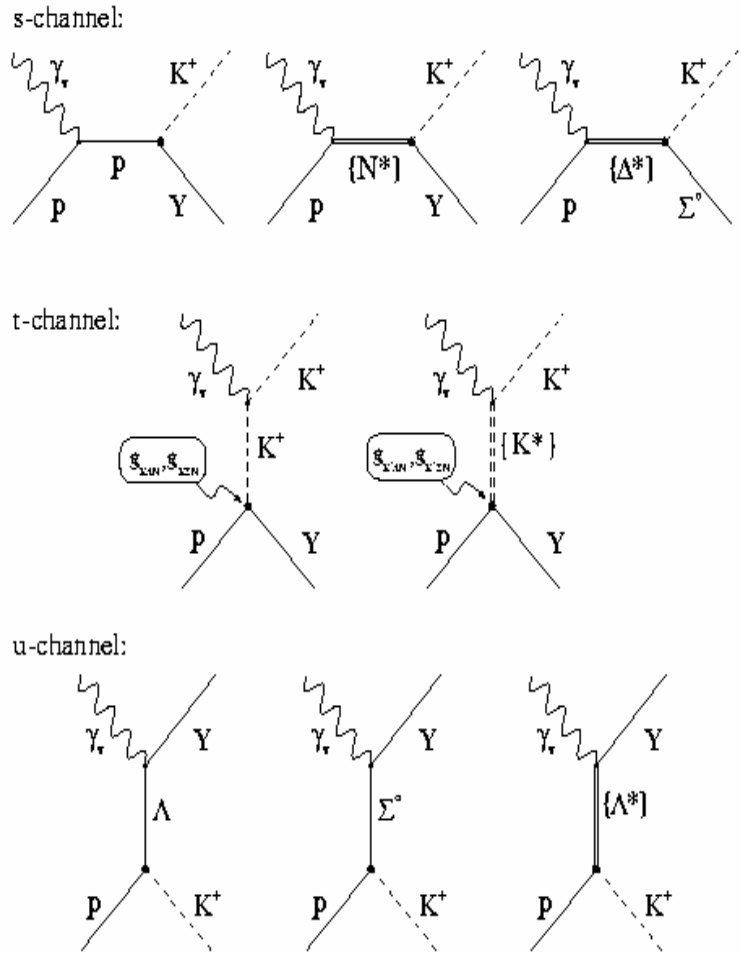

Fig. 2. Feynman diagrams for kaon electroproduction considered in an isobaric model

The kaon electromagnetic form factor has been extracted using the longitudinal component of the electroproduction cross section, $\sigma_L$ [1, 9, 14, 15]. The method generally used is the Chew-Low extrapolation technique. The contribution due to the pole term is parametrized as

$$\sigma_L \cong \frac{-2tQ^2}{(t-M_K^2)^2} g_{KN\Lambda}^2(t) F_K^2(Q^2,t)$$

where t is the squared four-momentum transfer, and $g_{KN\Lambda}$ is the coupling to the $KN\Lambda$ vertex.

In this analysis, we calculate the form factor of kaon for a value of $Q^2=2.35$ and W=(1.80, 1.85, 1.98, 2.08) from Jlab site (2001) (Table 1). First, we calculate the $\sigma_L$ and $\sigma_T$ (Table 2) by using Rosenbluth separation (Figs. 3 & 4), then by Chew-Low extrapolation technique calculate the form factor (Fig. 5). Our obtained form factor is compared with the previous values from four different approaches (Fig. 6).

Table 1. Kinematical data for L/T separation (Jlab site 2001)

| E | E' | $Q^2$ | W | $\theta_e$ | $\theta_k$ | $P_k$ | $t_{min}$ | $\varepsilon$ | $(d\sigma/d\sigma_K)_{CM}$ |
|---|---|---|---|---|---|---|---|---|---|
| GeV | GeV | $(GeV/c)^2$ | GeV | Deg. | Deg. | GeV/c | $(GeV/c)^2$ | | (nb/sr) |
| 5.755 | 3.18 | 2.35 | 1.8 | 20.43 | 22.66 | 1.741 | -0.9498 | 0.8071 | 152.8 |
| 4.232 | 1.688 | 2.35 | 1.8 | 32.99 | 18.59 | 1.741 | -0.9498 | 0.6078 | 144.8 |
| 3.395 | 0.8853 | 2.35 | 1.8 | 52.48 | 13.81 | 1.741 | -0.9498 | 0.3586 | 128.1 |
| 5.618 | 2.981 | 2.35 | 1.85 | 21.48 | 21.38 | 1.893 | -0.8562 | 0.7812 | 140.2 |
| 4.232 | 1.593 | 2.35 | 1.85 | 33.99 | 17.48 | 1.893 | -0.8562 | 0.579 | 126.9 |
| 3.395 | 0.788 | 2.35 | 1.85 | 55.89 | 12.5 | 1.893 | -0.8562 | 0.3134 | 121.7 |
| 5.618 | 2.719 | 2.35 | 1.98 | 22.51 | 18.83 | 2.266 | -0.6737 | 0.7368 | 135.9 |
| 4.506 | 1.634 | 2.35 | 1.98 | 32.82 | 15.78 | 2.266 | -0.6737 | 0.561 | 120 |
| 4.232 | 1.36 | 2.35 | 1.98 | 37.27 | 14.65 | 2.266 | -0.6737 | 0.4936 | 118.9 |
| 5.618 | 2.504 | 2.35 | 2.08 | 23.46 | 16.98 | 2.542 | -0.5716 | 0.6962 | 121.9 |
| 4.506 | 1.417 | 2.35 | 2.08 | 35.31 | 13.74 | 2.542 | -0.5716 | 0.4938 | 111.2 |
| 4.232 | 1.143 | 2.35 | 2.08 | 40.79 | 12.51 | 2.542 | -0.5716 | 0.4169 | 105 |

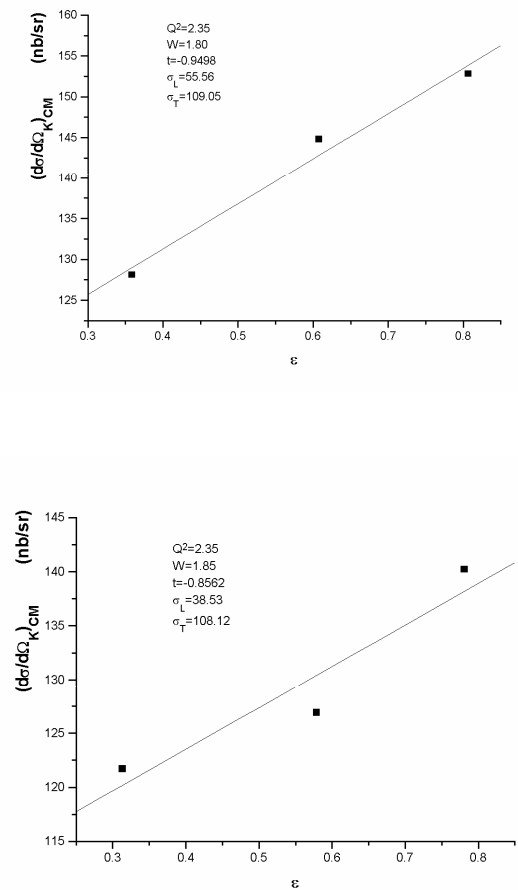

Fig. 3. Extraction of L/T separated cross section from Table 1

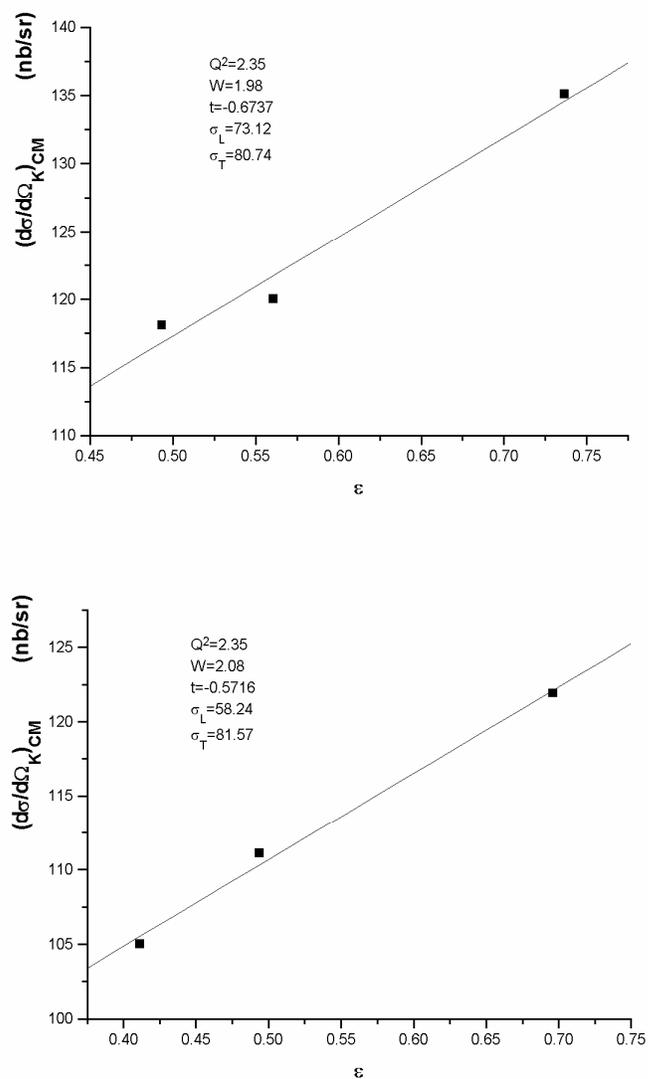

Fig. 4. Extraction of L/T separated cross section from Table 1

Table 2. L/T separation result from Figs. 3 & 4

| $Q^2$ | W | $\sigma_L$ | $\sigma_T$ |
|---|---|---|---|
| (GeV/c)$^2$ | GeV | (nb/sr) | (nb/sr) |
| 2.35 | 1.8 | 55.56 | 109.05 |
| 2.35 | 1.85 | 38.53 | 108.12 |
| 2.35 | 1.98 | 73.12 | 80.74 |
| 2.35 | 2.08 | 58.24 | 81.57 |

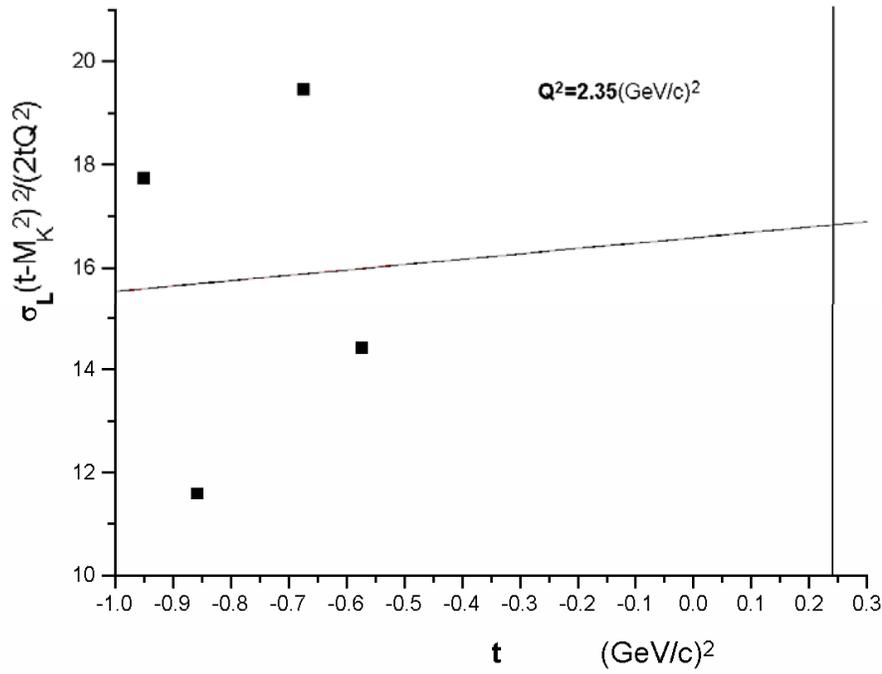

Fig. 5. Chew-Low extrapolation technique used for the extraction of the kaon form factor

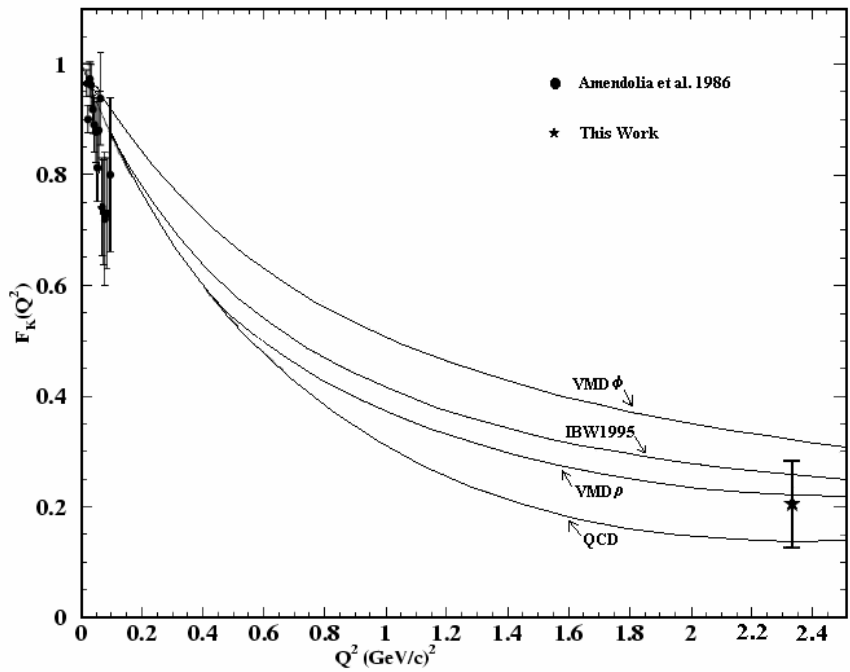

Fig. 6. Kaon form factor as a function of $Q^2$. The present extraction of $F_k(Q^2)$ ia shown at $Q^2=2.35$ $(GeV/c)^2$

## 4. CONCLUSIONS

The kaon form factor $F_K^+$ was extracted by extrapolating the measured t-dependence of the cross section to the kaon pole via a variation of the Chew-Low extrapolation technique. The t-dependence of the cross section was studied for the value of $Q^2$=2.35 (GeV/c)$^2$. For the form factor discussion, the available data is not enough, but this obtained form factor is in agreement with the previous works.

*Acknowledgements*- Partial support of Shiraz University research council is appreciated.